\documentclass[12pt,preprint]{aastex}

\shorttitle{Early Emission from GRB\,021211}
\shortauthors{Fox et al.}

\newcommand{\grb}{GRB\,021211}
\newcommand{\hete}{{\em HETE-2}}
\newcommand{\kshort}{\mbox{$K_{\rm s}$}}
\newcommand{\uJy}{\mbox{$\mu$Jy}}
\newcommand{\nuc}{\mbox{$\nu_c$}}
\newcommand{\nucs}{\mbox{$\nu_c^*$}}
\newcommand{\num}{\mbox{$\nu_m$}}
\newcommand{\nums}{\mbox{$\nu_m^*$}}

\newcommand{\nusa}{\mbox{$\nu_a$}}
\newcommand{\nuopt}{\mbox{$\nu_{\rm o}$}}
\newcommand{\nurad}{\mbox{$\nu_{\rm r}$}}
\newcommand{\tmopt}{\mbox{$t_{m,{\rm o}}$}}
\newcommand{\tmrad}{\mbox{$t_{m,{\rm r}}$}}
\newcommand{\Fmopt}{\mbox{$F_{m,{\rm o}}$}}
\newcommand{\Fmrad}{\mbox{$F_{m,{\rm r}}$}}
\newcommand{\Frad}{\mbox{$F_{\rm r}$}}
\newcommand{\pcmcube}{\mbox{cm$^{-3}$}}
\newcommand{\ergcmsq}{\mbox{erg cm$^{-2}$}}

\newcommand{\tpeak}{\mbox{$t_{\rm peak}$}}
\newcommand{\epse}{\mbox{$\epsilon_e$}}

\newcommand{\betaobs}{\mbox{$\beta_{\rm obs}$}}
\newcommand{\alphafs}{\mbox{$\alpha_{\rm fs}$}}
\newcommand{\alphars}{\mbox{$\alpha_{\rm rs}$}}
\newcommand{\alpharsr}{\alpha_{\rm rs}}

\newcommand{\simlt}{\lesssim}
\newcommand{\simgt}{\gtrsim}

\begin{document}

\title{Discovery of Early Optical Emission from GRB\,021211}

\author{D.~W. Fox\altaffilmark{1},
        P.~A.~Price\altaffilmark{2,3},
        A.~M.~Soderberg\altaffilmark{1},
        E.~Berger\altaffilmark{1},
        S.~R.~Kulkarni\altaffilmark{1},
        R.~Sari\altaffilmark{4},
        D.~A.~Frail\altaffilmark{5},
        F.~A.~Harrison\altaffilmark{2},
        S.~A.~Yost\altaffilmark{2},
        K.~Matthews\altaffilmark{1},
        B.~A.~Peterson\altaffilmark{3},
        I.~Tanaka\altaffilmark{6},
        J.~Christiansen\altaffilmark{3} 
        \&\
	G.~H.~Moriarty-Schieven\altaffilmark{7}
}
\altaffiltext{1}{Caltech Optical Observatories 105-24, 
       California Institute of Technology, Pasadena, CA 91125;
        \textit{derekfox@astro.caltech.edu, ams@astro.caltech.edu,
        ejb@astro.caltech.edu, srk@astro.caltech.edu,
        kym@caltech.edu}}
\altaffiltext{2}{Space Radiation Laboratory 220-47, California
        Institute of Technology, Pasadena, CA 91125;
        \textit{pap@srl.caltech.edu, fiona@srl.caltech.edu, 
        yost@srl.caltech.edu}} 
\altaffiltext{3}{Research School of Astronomy and Astrophysics, Mount
       Stromlo and Siding Spring Observatories, via Cotter Road,
       Weston Creek 2611, Australia; \textit{pap@mso.anu.edu.au,
        peterson@mso.anu.edu.au}}
\altaffiltext{4}{Theoretical Astrophysics 130-33, California Institute
        of Technology, Pasadena, CA 91125;
        \textit{sari@tapir.caltech.edu}} 
\altaffiltext{5}{National Radio Astronomy Observatory, Socorro, NM
        87801; \textit{dfrail@nrao.edu}}
\altaffiltext{6}{National Astronomical Observatory of Japan, Mitaka,
        Tokyo, 181-8588, Japan; \textit{itanaka@optik.mtk.nao.ac.jp}} 
\altaffiltext{7}{National Research Council of Canada, Joint Astronomy
        Centre, 660 N. A'ohoku Place, Hilo, HI 96720;
        \textit{g.moriarty-schieven@jach.hawaii.edu}}


\begin{abstract}

We report our discovery and early time optical, near-infrared, and
radio wavelength follow-up observations of the afterglow of the
gamma-ray burst \grb.  Our optical observations, beginning 21\,min
after the burst trigger, demonstrate that the early afterglow of this
burst is roughly three magnitudes fainter than the afterglow of
GRB\,990123 at similar epochs, and fainter than almost all known
afterglows at an epoch of 1\,d after the GRB.  Our near-infrared and
optical observations indicate that this is not due to extinction.
Combining our observations with data reported by other groups, we
identify the signature of a reverse shock.  This reverse shock is not
detected to a 3-$\sigma$ limit of 110\,\uJy\ in an 8.46-GHz VLA
observation at $t=0.10$\,d, implying either that the Lorentz factor of
the burst $\gamma\simlt 200$, or that synchrotron self-absorption
effects dominate the radio emission at this time.  Our early optical
observations, near the peak of the optical afterglow (forward shock),
allow us to characterize the afterglow in detail.  Comparing our model
to flux upper limits from the VLA at later times, $t\simgt 1$\,week,
we find that the late-time radio flux is suppressed by a factor of two
relative to the $\simgt$80\,\uJy\ peak flux at optical wavelengths.
This suppression is not likely to be due to synchrotron
self-absorption or an early jet break, and we suggest instead that the
burst may have suffered substantial radiative corrections.

\end{abstract}



\section{Introduction}
Gamma-ray burst (GRB) afterglow emission on timescales of hours to
days, the historical standard for ground-based follow-up observations,
can provide a robust measure of the global parameters of the GRB such
as its total explosive yield and the density of the circumburst medium
on length scales of $r\simgt 10^{17}$\,cm \citep{spn98,wg99}.  

Early emission, on a timescale of tens of minutes and less, can offer
insight into the details of the explosion, including the relativistic
Lorentz factor of the burst ejecta \citep{sp99a} and the distribution
of the circumburst medium on length scales $r\simgt 10^{15}$\,cm.  If
the GRB progenitor is a massive star, then this regime is expected to
show the clear signature of a dense stellar wind (density $\rho\propto
r^{-2}$; \citealt{cl00}).

These two unique diagnostics, among others, have motivated searches
for early optical emission from GRBs. The first success came with
detection of early optical \citep{abb+99} and radio \citep{kfs+99}
emission from GRB\,990123 \citep{bbk+99}. This emission has been
interpreted in terms of a reverse shock plowing back through the GRB
ejecta \citep{sp99b,mr99}. 

After nearly four years, early emission was recently discovered again
from GRB\,021004 \citep{fox+03} and \grb\ \citep{GCN.1731}. These
discoveries were made possible by the prompt alerts of the \hete\
satellite \citep{ricker+02}, and the wide field and high sensitivity
of the 48-inch Palomar Oschin Schmidt telescope (P48) equipped with
the NEAT camera \citep{prh+99}. The increase in the number of robotic
telescopes has allowed for intensive follow-up of both of these bursts
during the critical early minutes to hours.  Here, we report the
discovery of early optical emission and subsequent multi-wavelength
follow-up observations of \grb.


\section{Observations}
\label{sec:obs}
\begin{figure}[t]
\plotone{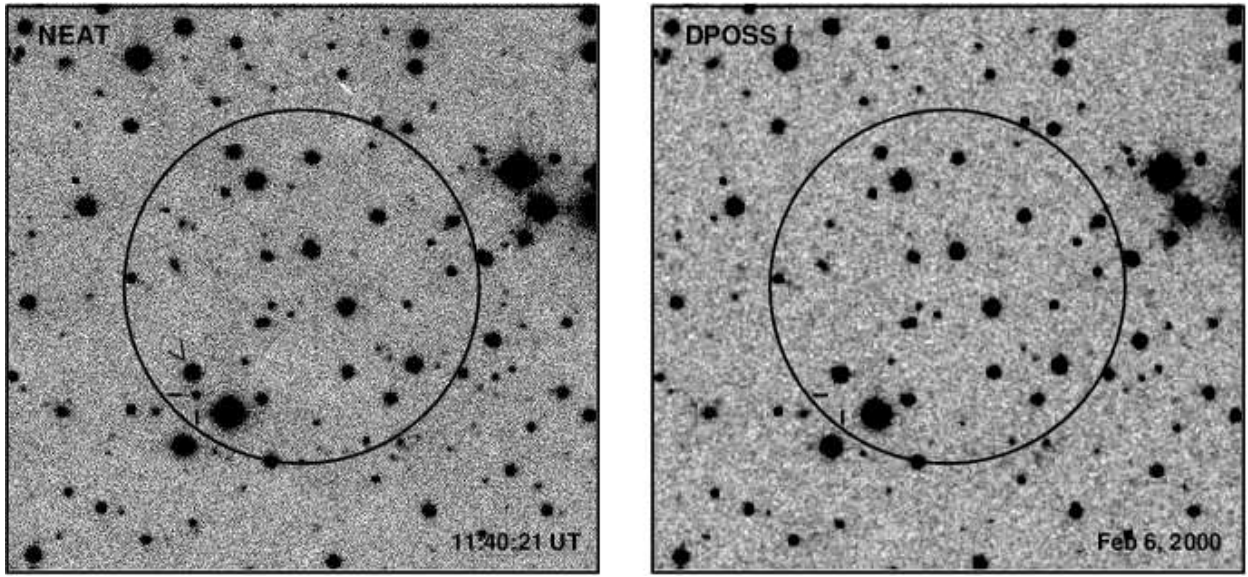}
\caption[]{\small %
Discovery image of the optical afterglow of \grb, as obtained with the
Near Earth Asteroid Tracking (NEAT) camera on the Oschin 48-inch
telescope, Palomar Observatory, 21\,min after the burst (left).  The
$R$-band (f-emulsion) Palomar Digital Sky Survey (DPOSS) image of the
GRB localization region is shown for comparison (right).  The
afterglow is marked by two dashes. The circle of radius 2\,arcminutes
marks the subsequent SXC localization.  The position of S1, one of our
secondary standards for NIR photometry (\S\ref{sec:obs}), is
indicated by a carat on the NEAT image; S2 is not visible in either
image.  Coordinates of S1 are $\alpha$=08:09:00.0, $\delta$=+06:43:52
(J2000) and coordinates of S2 are $\alpha$=08:09:00.3,
$\delta$=+06:43:38 (J2000).
}
\label{fig:discovery}
\end{figure}

\grb\ (HETE Trigger \#2493) was detected at UT 11:18:34 on 11 Dec
2002.  With a duration of $6$\,s and a fluence of $10^{-6}$\,erg
cm$^{-2}$ (8--40\,keV), the event is a typical long duration GRB
\citep{GCN.1734}.  At the time of electronic notification the \hete\
localization, although large (30-arcminute radius), was covered by the
1.1$^\circ\times 1.1^\circ$ field-of-view of NEAT. We identified a new
source that was not present in the Palomar Sky Survey
(Fig.~\ref{fig:discovery}) and announced this as a possible optical
counterpart of \grb\ \citep{GCN.1731}.  The best \hete\ localization
at the time of our discovery announcement was 14\,arcmin in radius; a
subsequent SXC localization produced the 2\,arcmin-radius circle shown
in Figure~\ref{fig:discovery}.  The position of the optical transient
(OT) with respect to the Guide Star Catalog~II is
$\alpha$=08:08:59.858 and $\delta$=+06:43:37.52 (J2000) with estimated
uncertainty of 0.1\,arcsec in each axis.

We continued monitoring with P48, the 40-inch telescope at Siding
Spring (SS0-40), and the Palomar Hale 5-m (P200) telescope with JCAM
\citep{bloom+03}.  Photometry has been performed using the secondary
calibrators of \citet{GCN.1753}; see Table~\ref{tab:photometry}.

Our optical source was subsequently identified in images obtained by
several robotic telescopes and other ground-based facilities. In
particular, RAPTOR \citep{GCN.1757}, KAIT \citep{GCN.1737} and S-LOTIS
\citep{GCN.1736} detected the source 90\,s, 108\,s, and 143\,s after the
burst, respectively.  In Figure~\ref{fig:MOAL} we present the light
curve obtained by combining data from these and other GCNs with our
data.  

{\it Near-infrared --}
We observed the afterglow in the near-infrared with the Cassegrain
f/70 low-background secondary and \mbox{D-78} infrared camera
\citep{soifer+91} on P200 on Dec.~11.55 UT (\kshort-band) and with
NIRC \citep{ms94} on the Keck-I 10-m telescope on Dec.~18.5 UT
($JH\kshort$).  NIRC observations were photometered relative to
SJ~9116 and SJ~9134 \citep{persson+98}, and the P200 zero-point was
determined from the NIRC observations by reference to two secondary
calibrators, S1 and S2 (see Fig.~\ref{fig:discovery}).  Our results
are given in Table~\ref{tab:photometry}.

{\it Radio --}
We searched for radio emission with the Very Large Array
(VLA)\footnotemark\footnotetext{The NRAO is a facility of the National
Science Foundation operated under cooperative agreement by Associated
Universities, Inc.}; our initial observation at $t=2.44$\,hr
represents the earliest radio observation of any GRB to date.  We used
PKS~0734+100 for phase calibration and 3C\,147 and 3C\,286 for flux
calibration.  No counterpart to the OT is detected in any single
observation.  Adding the data between 2002 December~20 and 2003
January~6, we measure an 8.46-GHz upper limit of 35\,\uJy\ at the
position of the OT.  Peak flux densities at the position of the OT can
be found in Table~\ref{tab:radio}.

Separately, on Dec~12.51 UT we carried out an observation in the
347-GHz band with the Submillimetre Common User Bolometer Array
(SCUBA) on the James Clark Maxwell Telescope
(JCMT)\footnotemark\footnotetext{The James Clerk Maxwell Telescope is
operated by The Joint Astronomy Centre on behalf of the Particle
Physics and Astronomy Research Council of the United Kingdom, the
Netherlands Organization for Scientific Research, and the National
Research Council of Canada.}.  We used CRL\,618 as the secondary flux
standard.  At the position of the OT we measure a 347-GHz flux density
of $2.0\pm 2.5$\,mJy.


\section{A Two-Component Light Curve}
\label{sec:Light-Curve}
\begin{figure}[t]
\epsscale{0.8}
\plotone{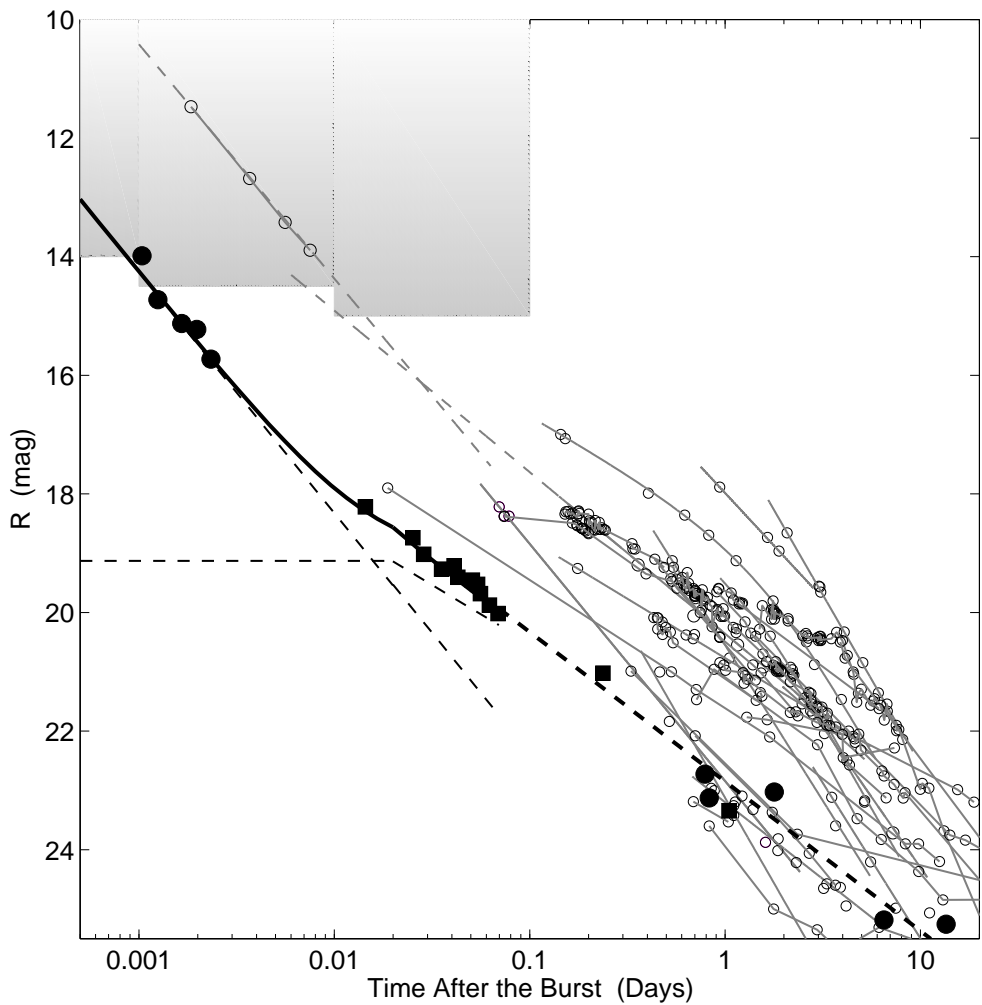}
\caption[]{\small %
Light curve of \grb\ and other GRBs, including a two-component flare +
afterglow fit for the early optical emission from \grb. The data for
\grb\ are drawn from the GCN literature (circles; see text) and this
work (squares; Table~\ref{tab:photometry}).  We fit the data at
$t<0.1$\,d only (see \S\ref{sec:aglow} for details); the dotted line
at $t>0.1$\,d is merely illustrative.  The last two points ($t>5$\,d)
are derived from HST observations \citep{GCN.1781} and should be
largely free of host contamination.  The data for other GRBs are drawn
from the literature (see \citealt{berger+02}).
}
\label{fig:MOAL}
\end{figure}

To date, only three bursts have had early-time optical detections:
GRB\,990123 \citep{abb+99}, GRB\,021004 \citep{fox+03} and
GRB\,021211.  \grb\ is at any epoch one of the faintest detected
afterglows; at early epochs it is more than 3\,mag fainter than
GRB\,990123, and at later epochs ($t\simgt 1$\,d) it lies below the
great majority of afterglow lightcurves (Fig.~\ref{fig:MOAL}).  As we
show below, this is not the result of extinction in the host galaxy,
so this burst reminds us that many so-called ``dark bursts'' may
simply be faint (c.f.\ \citealt{fynbo+01,berger+02}).

Examining Figure~\ref{fig:MOAL}, we see that the early behavior of
\grb\ bears a striking resemblance to the light curve of GRB\,990123
(c.f.\ \citealt{GCN.1754}).  The light curve can be considered as
composed of two segments: an initial steeply-declining ``flash'',
$f\propto t^{-\alpha}$ with $\alphars\approx 1.6$, followed by
emission declining as a typical afterglow, $\alphafs\approx 1$.  The
corresponding terms are ``flare'' and afterglow for the radio emission
\citep{kfs+99}.

The optical flash is thought to result from a reverse shock
propagating into the GRB ejecta \citep{sp99b}, while the afterglow
arises in the shocked ambient material swept up by the ejecta
\citep{spn98}. For each shock, it is assumed that some fraction of the
shock energy is partitioned into energetic electrons ($\epsilon_e$)
and magnetic fields ($\epsilon_B$). The electrons are expected to
follow a power-law distribution in energy, $dN/d\gamma \propto
\gamma^{-p}$ for $\gamma>\gamma_m$, where $\gamma$ is the electron
Lorentz factor. Their collective emission follows $f_\nu\propto
\nu^{-(p-1)/2}$ at frequencies above the synchrotron peak frequency
\num\ and $f_\nu\propto (\nu/\nu_m)^{1/3}$ otherwise.  Below the
self-absorption frequency \nusa, synchrotron self-absorption becomes
important, and above the characteristic cooling frequency \nuc,
electrons lose their energy rapidly, on the timescale of the shock
evolution.

The forward and reverse shocks evolve differently.  Once the reverse
shock has passed through the ejecta shell then no new electrons are
accelerated and the shocked electrons cool adiabatically:
$\nu_m\propto t^{-73/48}$, with the flux at $\nu_m$ (the ``peak''
flux) decaying as $F_{\nu_m}\propto t^{-47/48}$ \citep{sp99a}; for
$\nu>\nu_c$ the emission decreases even more rapidly. In contrast,
electrons are continually added to and accelerated at the forward
shock; in the case of a uniform circumburst medium, $F_{\nu_m}$ is
constant, and even above \nuc\ the flux declines only slowly.  Since
in a stellar wind-type circumburst medium ($\rho\sim r^{-2}$) the
reverse shock \nuc\ is inevitably below the optical band, detection of
long-lived reverse shock emission for \grb\ suffices to rule out wind
models for this burst \citep{cl00}.

\subsection{The Reverse Shock Flare}
\label{sec:flare}
The optical flux is already declining at the first epoch
\citep{GCN.1757}, when $t=t^*=90$\,s and $R_c=14$\,mag, corresponding
to $f=f^*=7.7$\,mJy at a mean frequency of $\nuopt=4.7\times
10^{14}$\,Hz; here the superscript * identifies quantities taken at
$t=90$\,s.  This implies that $\nums<\nuopt$, and constrains the time
when the flare spectral peak \num\ passes through the optical
bandpass, \tmopt, to be $\tmopt<t^*$.  Since the observed decline at
$t>t^*$ is not very rapid, $\alphars\simlt 2$, we have also that
$\nucs>\nuopt$.

The constraint on the time of the optical peak flux can be converted
into a constraint on the relativistic Lorentz factor $\gamma$ of the
GRB ejecta.  From \citet{sp99a} we have $\tpeak = 5\, n_0^{-1/3}
\gamma_{300}^{-8/3} E_{52}^{1/3}$\,s (thin-shell case), where \tpeak\
is the time at which the reverse shock crosses the GRB ejecta, $n_0$
is the circumburst particle density per cubic cm, $\gamma_{300}$ is
$\gamma/300$, and $E_{52}$ is the isotropic-equivalent burst energy in
units of $10^{52}$\,erg.  Since $\tpeak<\tmopt<t^*$, we find that
$\gamma > 100\, n_0^{-1/8} E_{52}^{1/8}$.

Using the prompt emission fluence of $10^{-6}$\,\ergcmsq\ (8--40\,keV)
together with the proposed redshift of $z=1.006$ \citep{GCN.1785}, we
find an isotropic $\gamma$-ray energy of $E_{\rm iso}=3\times
10^{51}$\,ergs for \grb.  Since this fluence is measured over a
relatively narrow band, the full $k$-corrected fluence \citep{bfs01}
may easily reach $10^{52}$\,ergs, and it is reasonable to assume
$E_{52}\sim 1$.

Following \citet{sp99b}, in the absence of synchrotron self-absorption
the radio flux will follow $\Frad\propto t^{-17/36}$ for times
$t<\tmrad$, and $\Frad\propto t^{-\alpharsr}$ for times $t>\tmrad$,
where \tmrad\ is the time when $\num$ passes through the radio band,
$\nurad=8.5$\,GHz.  The value of \tmrad\ follows directly from the
equation $\tmrad=\tmopt (\nuopt/\nurad)^{48/73}$, and is equal to
1.35\,d in the limiting case $\tmopt=t^*$.  By this time the peak flux
will have declined as $\Fmrad=\Fmopt (\tmrad/\tmopt)^{-47/48}$.  Note
that we are temporarily ignoring extinction effects, which are
discussed in \S\ref{sec:aglow} below.

For $\tmopt<t^*$, we extrapolate the observed optical flux decay back
in time as $\Fmopt=f^*(\tmopt/t^*)^{-\alpharsr}$.  The light curve of
the radio flare is thus a direct function of \tmopt:
\begin{equation}
\label{eq:fmrad}
   \Frad(t\le\tmrad) = 8.9\times 10^{-4} \, f^*
                       \left(\frac{\tmopt}{t^*}\right)^{-\alpharsr}
                       \left(\frac{t}{\tmrad}\right)^{-17/36},
\end{equation}
with the decay past \tmrad\ steepening to $t^{-\alpha_{\rm rs}}$.  Our
3-$\sigma$ upper limits at $t=0.10$\,d and $t=0.92$\,d are 110\,\uJy\
and 66\,\uJy, respectively.  For $\alphars\approx 1.6$
(\S\ref{sec:aglow}), the radio flux at $t=0.92$\,d is predicted to be
$\sim$15\,\uJy, independent of \tmopt.  The radio flux at $t=0.10$\,d
is a strong function of \tmopt, however, and Eq.~\ref{eq:fmrad} gives
$\tmopt>20$\,s and $\Fmopt<135$\,mJy.

A lower limit on \tmopt\ implies an upper limit on the Lorentz factor
of the ejecta, $\gamma < 180\, n_0^{-1/8} E_{52}^{1/8}$.  Following the
treatment of \citet{sp99b}, however, we find that synchrotron
self-absorption may be suppressing the radio flux below the level of
Eq.~\ref{eq:fmrad}, which would render this limit invalid.

\subsection{The Forward Shock Afterglow}
\label{sec:aglow}
A variety of afterglow lightcurves are possible \citep{kobayashi00}
but the form of sharp rise, plateau, and power-law decay is
near-universal. Given the bright flare, we cannot constrain the rising
portion of the afterglow lightcurve.  We fit the flare as a power-law
decay of index \alphars\ and the afterglow as a constant flux, $f_m$,
which undergoes a power-law decay with index \alphafs\ for $t>t_b$
(Fig.~\ref{fig:MOAL}).

We first examine the constraints on the reverse shock decay.
Published GCN data are sufficient to constrain $\alphars=1.63\pm
0.13$, and since $\alphars = \frac{47}{48} + \frac{73}{48}
(\frac{p-1}{2})$ \citep{sp99b}, this implies that $p=1.85\pm 0.17$.

The gap between the GCN data and our P48 measurements creates fitting
degeneracies in the full model, so we fix \alphars\ at three
successive values, 1.50, 1.63, and 1.76, and make fits to the full
dataset.  Incorporating the range in parameter uncertainties covered
in these three fits, we find that $\alphafs = 0.80\pm 0.13$,
$f_m=65\pm 20$\,\uJy\ and $t_b=1700\pm 280$\,s.  Note that because the
flare dominates the afterglow emission at early times, our fitted
$f_m$ should be considered a lower limit and our $t_b$ an upper limit.

We can measure the power-law slope of the spectrum, $\betaobs=-0.98$,
using our $B$ and \kshort\ photometry near $t=0.1$\,d (correcting
first for Galactic extinction of $A_R$=0.07\,mag; \citealt{sfd98}).
If $\nuc<\nuopt$ at this time, then $\alphafs=3/4(p-1)+1/4$, $p=1.7\pm
0.2$, and $\beta = -p/2 = -0.85\pm 0.1$, implying little extinction in
the host, $A_R=0.15\pm 0.18$ in the observer frame.  On the other
hand, if $\nuc>\nuopt$ at this time, then $\alphafs=3/4(p-1)$,
$p=2.1\pm 0.2$, and $\beta = -(p-1)/2 = -0.55\pm 0.1$, corresponding
to an intrinsic reddening of $A_R=0.48\pm 0.18$ (observer frame).  In
either case, extinction within the host cannot explain the
$\sim$2\,mag gap between the $R$-band lightcurve and those of most
afterglows.


Since we do not know the location of \nuc\ relative to the optical
band, we adopt an average extinction correction of $A_R=0.3$.  This
corrects our limiting peak flux of $f_m\simgt 65$\,\uJy\ to an
unextinguished flux of $F_m\simgt 85$\,\uJy.

From our model fits, $\alphafs=0.8\pm 0.13$, so that $F_m \approx 85
(\tmopt/t_b)^{-0.8}$\,\uJy.  We can use this relation along with Eqs.\
4.3 and 4.5 of \citet{se01} to estimate the circumburst density on
length scales of $r\sim 2\gamma^2 t_b c\simlt 10^{18}$\,cm.  We find
that $n_0 = 0.1 \epsilon_{B,-2}^{-23/15} \epsilon_{e,-1}^{-32/15}
E_{52}^{-38/15}$\,\pcmcube, where $\epsilon_e=0.1\epsilon_{e,-1}$.
Since we have $E_{52}\sim 1$ (\S\ref{sec:flare}), this implies either
a low-density circumburst medium, or abnormally low values of
$\epsilon_B$ or $\epsilon_e$.

In the absence of self-absorption, the radio afterglow can be expected
to peak at time $t\simlt t_b(\nuopt/\nurad)^{48/73} \simlt 25$\,d.
In the standard model the synchrotron peak flux $F_m$ remains constant
as the peak frequency decreases, so we expect a radio peak flux of
roughly 85\,\uJy.  Prior to the peak time, $\nurad<\num$ and
$\Frad\sim t^{0.5}$.  At the mean epoch of our summed image,
$t=17.3$\,d, we expect $\Frad\sim 70$\,\uJy, and our 2-$\sigma$ upper
limit is $\Frad<35$\,\uJy, a factor of two below the predicted level.

We can estimate the self-absorption frequency (Eq.~4.1 of
\citealt{se01}) in terms of the peak flux, $\nusa = 0.7\,
E_{52}^{-9/20}\, n_0^{3/20}\, (F_m/85\,\uJy)^{1.86}$\,GHz.
Self-absorption will be effective in the case of a high circumburst
density -- unlikely on the basis of our $n_0$ estimate -- or if $F_m$
is sufficiently large.  In the latter case, at fixed frequency the
effects of increasing self-absorption win out over the raw increase in
flux.  Suppressing the radio flux by a factor of two requires
$F_m\simgt800$\,\uJy\ ($\tmopt\simlt 100$\,s), however, which seems
unlikely.

The discrepancy between the observed and expected fluxes can be due to
an early jet break, $t_{\rm jet}<\tmrad$ \citep{rhoads99,sph99}.  The
optical light-curve, however, shows no sharp steepening out to
$t\simgt 10$\,d (Fig.~\ref{fig:MOAL}).  Moreover, the measured fluence
and proposed redshift $z=1.006$ for \grb\ are consistent with minimal
beaming for a ``standard candle'' burst ($E_{\rm iso}\sim 5\times
10^{50}$\,ergs; \citealt{frail+01}), so that an early jet-break seems
an unlikely scenario to explain the suppression of the radio flux.

Alternatively, the afterglow radiative corrections
\citep{sari97,cps98} may be significant.  In the slow-cooling regime
($\nuc>\num$), radiative corrections can be substantial if \epse\ is
large ($\epse\simgt 0.1$), \nuc\ is close to \num, or $p$ is close or
equal to 2.  We have not derived an estimate for \epse, but $p\approx
2$ for our models, so this scenario remains an attractive candidate
for future modeling.


\section*{Acknowledgments.}
We acknowledge the efforts of the NEAT team at JPL, and of Scott
Barthelmy at Goddard for the GCN.  Thanks are due to Josh Bloom, who
has built an excellent transient observing system in JCAM.  SRK thanks
S.~Thorsett for maintaining the ``hyperlinked GCN'' web page.  GRB
research at Caltech is supported by grants from NSF and NASA.


\begin{thebibliography}{36}
\expandafter\ifx\csname natexlab\endcsname\relax\def\natexlab#1{#1}\fi

\bibitem[{{Akerlof} {et~al.}(1999){Akerlof}, {Balsano}, {Barthelemy}, {Bloch},
  {Butterworth}, {Casperson}, {Cline}, {Fletcher}, {Frontera}, {Gisler},
  {Heise}, {Hills}, {Kehoe}, {Lee}, {Marshall}, {McKay}, {Miller}, {Piro},
  {Priedhorsky}, {Szymanski}, \& {Wren}}]{abb+99}
{Akerlof}, C. {et al.}\  1999, Nature, 398, 400

\bibitem[{{Berger} {et~al.}(2002){Berger}, {Kulkarni}, {Bloom}, {Price}, {Fox},
  {Frail}, {Axelrod}, {Chevalier}, {Colbert}, {Costa}, {Djorgovski},
  {Frontera}, {Galama}, {Halpern}, {Harrison}, {Holtzman}, {Hurley}, {Kimble},
  {McCarthy}, {Piro}, {Reichart}, {Ricker}, {Sari}, {Schmidt}, {Wheeler},
  {Vanderspek}, \& {Yost}}]{berger+02}
{Berger}, E. {et al.}\  2002, \apj, 581, 981

\bibitem[{{Bloom} {et~al.}(2001){Bloom}, {Frail}, \& {Sari}}]{bfs01}
{Bloom}, J.~S., {Frail}, D.~A., \& {Sari}, R. 2001, \aj, 121, 2879

\bibitem[{{Bloom} {et~al.}(2003){Bloom}, {Kulkarni}, {Clemens}, {Diercks},
  {Simcoe}, \& {Behr}}]{bloom+03}
{Bloom}, J.~S., {Kulkarni}, S.~R., {Clemens}, J.~C., {Diercks}, A., {Simcoe},
  R.~A., \& {Behr}, B.~B. 2003, Publ. Astr. Soc. Pacific, submitted

\bibitem[{{Briggs} {et~al.}(1999){Briggs}, {Band}, {Kippen}, {Preece},
  {Kouveliotou}, {van Paradijs}, {Share}, {Murphy}, {Matz}, {Connors},
  {Winkler}, {McConnell}, {Ryan}, {Williams}, {Young}, {Dingus}, {Catelli}, \&
  {Wijers}}]{bbk+99}
{Briggs}, M.~S. {et al.}\  1999, Astrophys. J., 524, 82

\bibitem[{{Chevalier} \& {Li}(2000)}]{cl00}
{Chevalier}, R.~A. \& {Li}, Z. 2000, Astrophys. J., 536, 195

\bibitem[{{Chornock} {et~al.}(2002){Chornock}, {Li}, {Filippenko}, \&
  {Jha}}]{GCN.1754}
{Chornock}, R., {Li}, W., {Filippenko}, A.~V., \& {Jha}, S. 2002, GRB Circular
  Network, 1754, 1

\bibitem[{{Cohen} {et~al.}(1998){Cohen}, {Piran}, \& {Sari}}]{cps98}
{Cohen}, E., {Piran}, T., \& {Sari}, R. 1998, \apj, 509, 717

\bibitem[{{Crew} {et~al.}(2002){Crew}, {Villasenor}, {Vanderspek}, {Doty},
  {Monnelly}, {Butler}, {Cline}, {Jernigan}, {Levine}, {Martel}, {Morgan},
  {Prigozhin}, {Azzibrouck}, {Braga}, {Manchanda}, {Pizzichini}, {Ricker},
  {Atteia}, {Kawai}, {Lamb}, {Woosley}, {Shirasaki}, {Graziani}, {Matsuoka},
  {Tamagawa}, {Torii}, {Sakamoto}, {Yoshida}, {Fenimore}, {Galassi},
  {Tavenner}, {Donaghy}, {Nakagawa}, {Takahashi}, {Suzuki}, {Satoh}, {Urata},
  {Boer}, {Olive}, {Dezalay}, {Barraud}, \& {Hurley}}]{GCN.1734}
{Crew}, G. {et al.}\  2002, GRB Circular Network, 1734, 1

\bibitem[{{Fox} \& {Price}(2002)}]{GCN.1731}
{Fox}, D.~W. \& {Price}, P.~A. 2002, GRB Circular Network, 1731, 1

\bibitem[{{Fox} {et~al.}(2003){Fox}, {Yost}, {Kulkarni}, {Frail}, {Torri},
  {Kato}, {Yamaoka}, {Sako}, {Harrison}, {Sari}, {Price}, {Berger},
  {Soderberg}, {Djorgovski}, {Barth}, {Pravdo}, {Frail}, {Gal-Yam}, {Lipkin},
  {Mauch}, {Harrison}, \& {Buttery}}]{fox+03}
{Fox}, D.~W. {et al.}\  2003, Nature, submitted

\bibitem[{{Frail} {et~al.}(2001){Frail}, {Kulkarni}, {Sari}, {Djorgovski},
  {Bloom}, {Galama}, {Reichart}, {Berger}, {Harrison}, {Price}, {Yost},
  {Diercks}, {Goodrich}, \& {Chaffee}}]{frail+01}
{Frail}, D.~A. {et al.}\  2001, \apjl, 562, L55

\bibitem[{{Fruchter} {et~al.}(2002){Fruchter}, {Levan}, {Vreeswijk}, {Holland},
  \& {Kouveliotou}}]{GCN.1781}
{Fruchter}, A., {Levan}, A., {Vreeswijk}, P., {Holland}, S.~T., \&
  {Kouveliotou}, C. 2002, GRB Circular Network, 1781, 1

\bibitem[{{Fynbo} {et~al.}(2001){Fynbo}, {Jensen}, {Gorosabel}, {Hjorth},
  {Pedersen}, {M{\o}ller}, {Abbott}, {Castro-Tirado}, {Delgado}, {Greiner},
  {Henden}, {Magazz{\` u}}, {Masetti}, {Merlino}, {Masegosa}, {{\O}stensen},
  {Palazzi}, {Pian}, {Schwarz}, {Cline}, {Guidorzi}, {Goldsten}, {Hurley},
  {Mazets}, {McClanahan}, {Montanari}, {Starr}, \& {Trombka}}]{fynbo+01}
{Fynbo}, J.~U. {et al.}\  2001, \aap, 369, 373

\bibitem[{{Henden}(2002)}]{GCN.1753}
{Henden}, A. 2002, GRB Circular Network, 1753, 1

\bibitem[{{Kobayashi}(2000)}]{kobayashi00}
{Kobayashi}, S. 2000, \apj, 545, 807

\bibitem[{{Kulkarni} {et~al.}(1999){Kulkarni}, {Frail}, {Sari},
  {Moriarty-Schieven}, {Shepherd}, {Udomprasert}, {Readhead}, {Bloom},
  {Feroci}, \& {Costa}}]{kfs+99}
{Kulkarni}, S.~R. {et al.}\  1999, Astrophys. J., 522, L97

\bibitem[{{Li} {et~al.}(2002){Li}, {Filippenko}, {Chornock}, \&
  {Jha}}]{GCN.1737}
{Li}, W., {Filippenko}, A.~V., {Chornock}, R., \& {Jha}, S. 2002, GRB Circular
  Network, 1737, 1

\bibitem[{{M{\' e}sz{\' a}ros} \& {Rees}(1999)}]{mr99}
{M{\' e}sz{\' a}ros}, P. \& {Rees}, M.~J. 1999, Mon. Not. R. astr. Soc., 306,
  L39

\bibitem[{{Matthews} \& {Soifer}(1994)}]{ms94}
{Matthews}, K. \& {Soifer}, B.~T. 1994, Experimental Astronomy, 3, 77

\bibitem[{{Park} {et~al.}(2002){Park}, {Williams}, \& {Barthelmy}}]{GCN.1736}
{Park}, H.~S., {Williams}, G., \& {Barthelmy}, S. 2002, GRB Circular Network,
  1736, 1

\bibitem[{{Persson} {et~al.}(1998){Persson}, {Murphy}, {Krzeminski}, {Roth}, \&
  {Rieke}}]{persson+98}
{Persson}, S.~E., {Murphy}, D.~C., {Krzeminski}, W., {Roth}, M., \& {Rieke},
  M.~J. 1998, \aj, 116, 2475

\bibitem[{{Pravdo} {et~al.}(1999){Pravdo}, {Rabinowitz}, {Helin}, {Lawrence},
  {Bambery}, {Clark}, {Groom}, {Levin}, {Lorre}, {Shaklan}, {Kervin},
  {Africano}, {Sydney}, \& {Soohoo}}]{prh+99}
{Pravdo}, S.~H. {et al.}\  1999, Astron. J., 117, 1616

\bibitem[{{Rhoads}(1999)}]{rhoads99}
{Rhoads}, J.~E. 1999, \apj, 525, 737

\bibitem[{{Ricker} {et~al.}(2002){Ricker}, {Hurley}, {Lamb}, {Woosley},
  {Atteia}, {Kawai}, {Vanderspek}, {Crew}, {Doty}, {Villasenor}, {Prigozhin},
  {Monnelly}, {Butler}, {Matsuoka}, {Shirasaki}, {Tamagawa}, {Torii},
  {Sakamoto}, {Yoshida}, {Fenimore}, {Galassi}, {Tavenner}, {Donaghy},
  {Graziani}, {Boer}, {Dezalay}, {Niel}, {Olive}, {Vedrenne}, {Cline},
  {Jernigan}, {Levine}, {Martel}, {Morgan}, {Braga}, {Manchanda}, {Pizzichini},
  {Takagishi}, \& {Yamauchi}}]{ricker+02}
{Ricker}, G. {et al.}\  2002, \apjl, 571, L127

\bibitem[{{Sari}(1997)}]{sari97}
{Sari}, R. 1997, \apjl, 489, L37+

\bibitem[{{Sari} \& {Esin}(2001)}]{se01}
{Sari}, R. \& {Esin}, A.~A. 2001, \apj, 548, 787

\bibitem[{{Sari} \& {Piran}(1999{\natexlab{a}})}]{sp99b}
{Sari}, R. \& {Piran}, T. 1999{\natexlab{a}}, Astrophys. J., 517, L109

\bibitem[{{Sari} \& {Piran}(1999{\natexlab{b}})}]{sp99a}
--- 1999{\natexlab{b}}, Astrophys. J., 520, 641

\bibitem[{{Sari} {et~al.}(1999){Sari}, {Piran}, \& {Halpern}}]{sph99}
{Sari}, R., {Piran}, T., \& {Halpern}, J.~P. 1999, \apjl, 519, L17

\bibitem[{{Sari} {et~al.}(1998){Sari}, {Piran}, \& {Narayan}}]{spn98}
{Sari}, R., {Piran}, T., \& {Narayan}, R. 1998, \apjl, 497, L17+

\bibitem[{{Schlegel} {et~al.}(1998){Schlegel}, {Finkbeiner}, \&
  {Davis}}]{sfd98}
{Schlegel}, D.~J., {Finkbeiner}, D.~P., \& {Davis}, M. 1998, \apj, 500, 525

\bibitem[{{Soifer} {et~al.}(1991){Soifer}, {Neugebauer}, {Graham}, {Matthews},
  {Mazzarella}, {Lonsdale}, {Rowan-Robinson}, {Broadhurst}, {Lawrence}, \&
  {McMahon}}]{soifer+91}
{Soifer}, B.~T. {et al.}\  1991, \apjl, 381, L55

\bibitem[{{Vreeswijk} {et~al.}(2002){Vreeswijk}, {Fruchter}, {Hjorth}, \&
  {Kouveliotou}}]{GCN.1785}
{Vreeswijk}, P., {Fruchter}, A., {Hjorth}, J., \& {Kouveliotou}, C. 2002, GRB
  Circular Network, 1785, 1

\bibitem[{{Wijers} \& {Galama}(1999)}]{wg99}
{Wijers}, R.~A.~M.~J. \& {Galama}, T.~J. 1999, \apj, 523, 177

\bibitem[{{Wozniak} {et~al.}(2002){Wozniak}, {Vestrand}, {Starr}, {Wren},
  {Borozdin}, {Brumby}, {Casperson}, {Galassi}, {McGowan}, \&
  {White}}]{GCN.1757}
{Wozniak}, P. {et al.}\  2002, GRB Circular Network, 1757, 1

\end{thebibliography}

\clearpage


\begin{deluxetable}{llcccc}
\tablewidth{0pt}
\tablecaption{Optical Photometry for GRB021211}
\tablehead{
\colhead{Date}     &\colhead{$\Delta T$} & \colhead{Telescope}      &
\colhead{Filter}          & \colhead{Magnitude}  &
\colhead{Error} \\
\colhead{(UT)} & \colhead{(days)} & & & &}
\startdata

Dec 11.4854  & 0.0144     &   P48 & R$^a$ & 18.293  & 0.024  \\
Dec 11.4962  & 0.0252     &   P48 & R$^a$ & 18.813  & 0.045  \\
Dec 11.4997  & 0.0287     &   P48 & R$^a$ & 19.093  & 0.059  \\
Dec 11.5064  & 0.0354     &   P48 & R$^a$ & 19.343  & 0.071  \\
Dec 11.5120  & 0.0410     &   P48 & R$^a$ & 19.286  & 0.068  \\
Dec 11.5139  & 0.0429     &   P48 & R$^a$ & 19.479  & 0.077  \\
Dec 11.5218  & 0.0508     &   P48 & R$^a$ & 19.529  & 0.082  \\
Dec 11.5251  & 0.0541     &   P48 & R$^a$ & 19.598  & 0.086  \\
Dec 11.5270  & 0.0560     &   P48 & R$^a$ & 19.755  & 0.102  \\
Dec 11.5332  & 0.0622     &   P48 & R$^a$ & 19.950  & 0.110  \\
Dec 11.5399  & 0.0689     &   P48 & R$^a$ & 20.091  & 0.130  \\	
Dec 11.5533  & 0.0823     &   P200/IRcam &  Ks & 18.01 & 0.150 \\
Dec 11.6025  & 0.1310     &   SSO40/WFI & B & 21.877 & 0.138 \\
Dec 11.7083  & 0.2370     &   SSO40/WFI & R & 21.096 & 0.128 \\
Dec 12.518   & 1.037      &   P200/JCAM & g' &   23.398 & 0.047 \\
Dec 12.518   & 1.037      &   P200/JCAM & r' &   23.416 & 0.073 \\
Dec 18.49    & 7.019      &   K-I/NIRC & Ks & 22.12 & 0.18 \\
Dec 18.52    & 7.049      &   K-I/NIRC & J  & 22.77  & 0.11 \\
Dec 18.56    & 7.089      &   K-I/NIRC & H  & 22.48  & 0.16 \\
\enddata
\label{tab:photometry}
\tablenotetext{a}{P48 unfiltered observations have been photometered
  against the $R$-band calibration of \citet{GCN.1753}. }
\end{deluxetable}

\begin{deluxetable}{llccrr}
\tablewidth{0pt}
\tablecaption{Radio Observations of GRB\,021211}
\tablehead{
\colhead{Date} &
\colhead{$\Delta T$} & 
\colhead{Telescope}  & 
\colhead{Frequency}  & 
\colhead{Flux}  &
\colhead{Flux Error} \\
\colhead{(UT)}   & 
\colhead{(days)} &
\colhead{}       &
\colhead{(GHz)}  & 
\colhead{($\mu$Jy)} & 
\colhead{($\mu$Jy)}}
\startdata
2002 Dec 11.57 & 0.10 & VLA & 8.46 & 12 & 36\\
2002 Dec 12.39 & 0.92 & VLA & 8.46 & $-$36 & 22\\
2002 Dec 13.40 & 1.93 & VLA & 8.46 & 9 & 45\\
2002 Dec 13.40 & 1.93 & VLA & 22.5 & 48 & 62\\
2002 Dec 15.32 & 3.85 & VLA & 8.46 & 45 & 23\\
2002 Dec 16.29 & 4.82 & VLA & 8.46 & $-$6.5 & 19\\
2002 Dec 20.32 & 8.85 & VLA & 8.46  & 60 & 28\\
2002 Dec 22.26 & 10.79 & VLA & 8.46 & 0.3 & 27\\
2002 Dec 26.44 & 14.97 & VLA & 8.46 & $-$7.3 & 23\\
2002 Dec 28.55 & 17.08 & VLA & 8.46 & 33 & 28\\
2003 Jan 04.39 & 23.92 & VLA & 8.46 & 2.4 & 24\\
2003 Jan 06.28 & 25.81 & VLA & 8.46 & 46 & 24\\
\hline
Dec 20--Jan 06  & \omit & VLA & 8.46 & 15 & 10 \\
\enddata
\label{tab:radio}
\end{deluxetable}



\clearpage


\end{document}